\documentclass[manuscript]{aastex}
\usepackage{natbib,graphics}

\newcommand{\kms}{\mbox{km~s$^{-1}$}}
\newcommand{\tscl}{\mbox{$^{37}$Cl}}
\newcommand{\tfcl}{\mbox{$^{35}$Cl}}

\newcommand{\mh}{\mbox{H$_2$}}

\shorttitle{An HCl Survey}
\shortauthors{Peng et al.}

\begin{document}

\title{A Comprehensive Survey of Hydrogen Chloride in the Galaxy}

\author{Ruisheng Peng, Hiroshige Yoshida, and Richard A. Chamberlin}
\affil{Caltech Submillimeter Observatory, 111 Nowelo St., Hilo, HI 96720}
\email{peng@submm.caltech.edu, hiro@submm.caltech.edu, cham@astro.caltech.edu}

\author{Thomas G. Phillips and Dariusz C. Lis}
\affil{Division of Physics, Mathematics, and Astronomy, 320-47,
  California Institute of Technology, Pasadena, CA 91125} 
\email{tgp@submm.caltech.edu, dcl@submm.caltech.edu}

\and

\author{Maryvonne Gerin}
\affil{LERMA-LRA, CNRS, Observatoire de Paris and Ecole Normale Sup'erieure, 24 Rue Lhomond, F 75231 Paris Cedex 05, France}
\email{gerin@lra.ens.fr}

\begin{abstract}
  We report new observations of the fundamental $J=1-0$ transition of
  HCl (at 625.918GHz) toward a sample of 25 galactic star-forming
  regions, molecular clouds, and evolved stars, carried out using the
  Caltech Submillimeter Observatory. Fourteen sources in the sample
  are also observed in the corresponding H\tscl\ $J=1-0$ transition
  (at 624.978GHz). We have obtained clear detections in all but four
  of the targets, often in emission. Absorptions against
  bright background continuum sources are also seen in nine cases, usually
  involving a delicate balance between emission and absorption
  features. From RADEX modeling, we derive gas densities and HCl column
  densities for sources with HCl emission. HCl is found in a wide
  range of environments, with gas densities ranging from $10^5$ to
  $10^7$~cm$^{-3}$. The HCl abundance relative to H$_2$ is in the
  range of $(3-30)\times10^{-10}$. Comparing with the chlorine
  abundance in the solar neighborhood, this corresponds to a chlorine
  depletion factor of up to $\sim$400, assuming that HCl accounts for
  one third of the total chlorine in the gas phase. The [\tfcl]/[\tscl]
  isotopic ratio is rather varied, from unity to $\sim$5, mostly lower than
  the terrestrial value of 3.1. Such variation is highly localized,
  and could be generated by the nucleosynthesis in supernovae, which predicts
  a \tscl\ deficiency in most models. The lower ratios seen in W3IRS4 and 
  W3IRS5 likely confine the progenitors of the supernovae to stars with 
  relatively large mass ($\ga$25M$_\sun$) and high metallicity (Z$\sim$0.02).

\end{abstract}


\section{Introduction}

Initially of interest as a potential coolant for molecular clouds,
hydrogen chloride (HCl) was first detected in the $J=1-0$ transition
at 625.918~GHz \citep{delucia1971} toward IRc2 in Orion A, in emission, 
by \citet{bkp1985}, and in absorption against dust continuum in Sgr~B2, by
\citet{jonas1995}, both aboard the NASA Kuiper Airborne Observatory
with a beam width of 2\farcm3. Ground-based observations at a higher
spatial resolution of 12\arcsec\ were carried out using the 10.4-m
telescope of the Caltech Submillimeter Observatory (CSO) by
\citet{spw1995} and \citet{sfl1996} toward OMC-1 and MonR2.
\citet{sfl1996} also detected the H\tscl\ $J=1-0$ transition toward
OMC-1. These studies found the HCl abundance to be $10^{-9}-10^{-10}$
relative to molecular hydrogen, indicating that chlorine is depleted
by a factor of several hundred compared to its solar neighborhood
abundance, and that HCl is not a significant coolant of molecular
clouds. \citet{sfl1996} also reported a [\tfcl]/[\tscl] isotopic ratio
that is $1.3-2$ times higher than the solar neighborhood value,
indicating variations of the isotropic ratio in the Galactic
interstellar medium.

Chemistry of chlorine and hydrogen chloride in the interstellar medium has
been investigated by a number of authors
\citep{jura1974,dalgarno1974,bah1986,spw1995,nw2009}, establishing a
relatively simple chemical network involving less than a dozen other
atomic and molecular species, as well as electrons and cosmic-ray
particles, in the production and destruction of HCl. In dense molecular
clouds, such a network yields, among other species, predominantly HCl, taking 
up about one third of the chlorine in gas phase \citep{spw1995,nw2009}.

HCl $J=1-0$ transition falls on the shoulder of a relatively strong atmospheric
water absorption line. Consequently, ground-based observations of HCl
have been limited to the rare occasions of superb weather at high sites,
such as Mauna Kea, and relatively little is known about HCl in
different categories of objects, such as evolved stars, active star
forming regions, molecular clouds with no apparent star formation
activity, etc. In particular, the [$^{35}$Cl]/[$^{37}$Cl] isotopic
ratio is of interest in different categories of objects, as
variations of this ratio can be produced through nucleosynthesis in
massive stars followed by mixing of the processed material into the
surrounding molecular cloud \citep{sfl1996}. The chlorine isotopic
ratio could thus offer insights into the evolutionary history of the
molecular material.

To address these questions, we took advantage of several periods of 
superb weather on Mauna Kea to conduct a comprehensive survey of HCl in 
a sample of well known objects. In the following sections, we first describe 
the observational details, then present the results and data analysis, 
followed by discussions and conclusions.

\section{Observations}

Observations described here were carried out on September 11, 2008,
March 2, and April 7--15, 2010 with the 10.4~meter CSO telescope. The
690~GHz facility heterodyne receiver was tuned to 625.918756 GHz and
624.977821 GHz, respectively, for observations of HCl $J=1-0$ and
H\tscl\ $ J=1-0$ lines. The lines fall on the shoulder of a prominent
atmospheric water absorption line, therefore requiring premium weather
conditions. The atmospheric opacity measured at 225 GHz was mostly
0.03--0.04, and the typical double-side-band system temperature was 
2500~K. There were marginal periods in April 7-15, 2010 when atmospheric 
opacity was around 0.05--0.06, resulting in a typical system temperature of
6000~K. Both the FFTS and 4~GHz AOS facility spectrometers were used
in 1~GHz mode as backends.

The Dish Surface Optimization System (DSOS) \citep{leong2006} was
used to correct gravitational deformations of the telescope surface,
maintaining a clean 13\farcs5 beam over a wide range of elevations.
Temperature calibration using a hot chopper wheel yields $T_A^*$,
the antenna temperature corrected for atmosphere and telescope hot
spillover. A main beam efficiency of $\sim$50\% measured on Jupiter and
Uranus is used to convert $T_A^*$ to $T_{mb}$, the main beam
brightness temperature. Telescope pointing was checked on Jupiter,
Mars, and Uranus, and was found to be within 4\arcsec.

Observations were conducted in the beam switching mode at a frequency
of 0.523~Hz and with a chop throw of 300\arcsec. Typical integration
time on each source was 300 seconds, resulting in a rms of $\sim$0.15
K at a velocity resolution of 0.31 \kms. The source selection was
primarily based on its accessibility at reasonable elevations, with a
secondary goal of covering a diverse collection of objects. Due to its
high critical density ($\sim$10$^8$ cm$^{-3}$), the excitation of
the HCl $J=1-0$ transition is most likely sub-thermal, and radiative
excitation, in the presence of continuum emission, is likely to play 
an important role in the excitation of the transition. The presence 
of dense gas and/or continuum radiation in an object would likely
yield a detection. In the end, 25 sources/positions were
observed, among them active star forming regions, giant molecular
clouds with no apparent star forming activities, evolved stars,
shocked gas in a SNR, and ultra-compact HII regions.
Table~\ref{tab:obs-sum} lists observational details for all observed
sources.

\section{Results}

\begin{figure}
\plotone{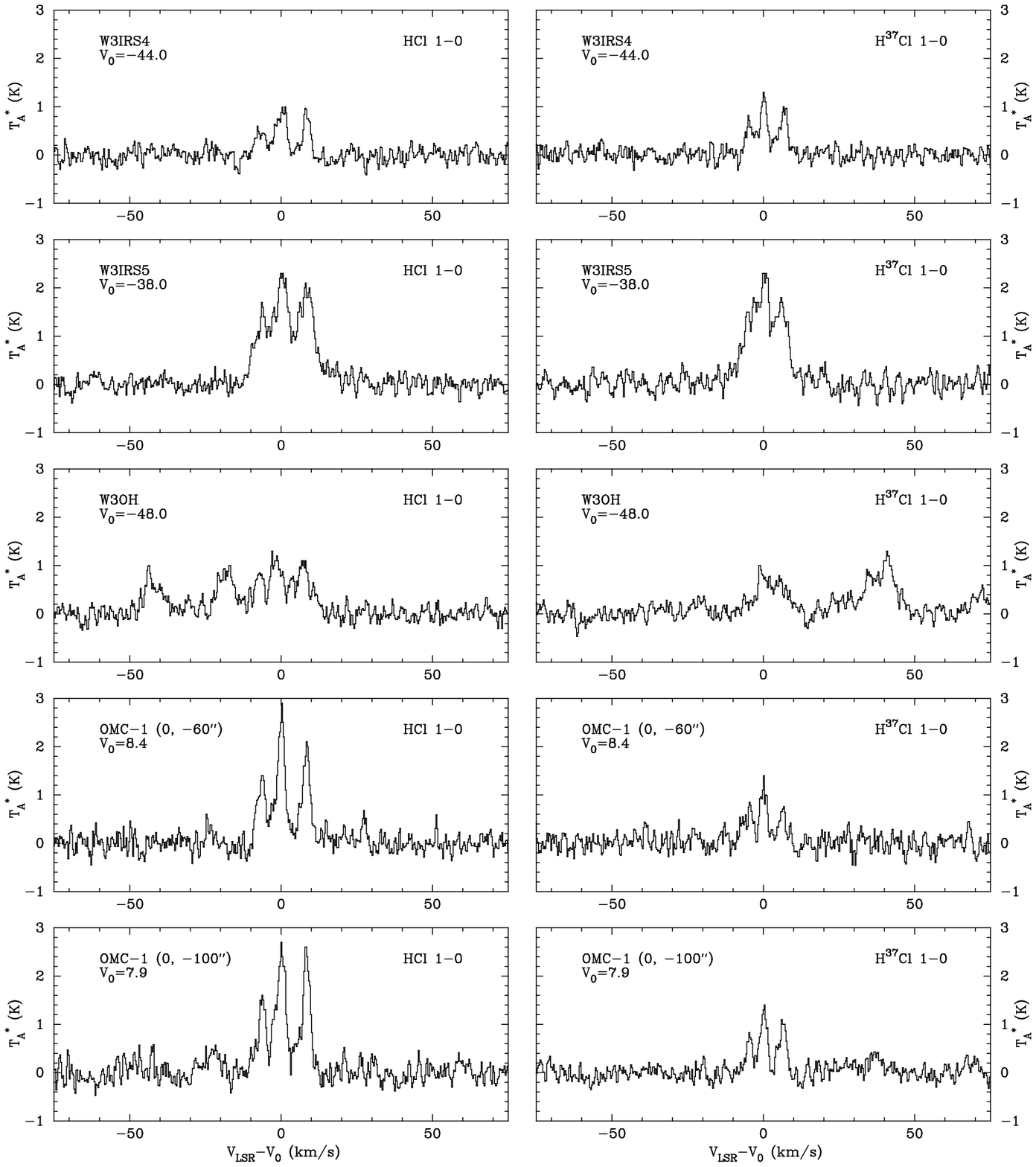}
\caption{HCl and H\tscl\  $J=1-0$ spectra toward W3IRS4, W3IRS5, W3(OH), and two
positions in OMC-1 offset from IRc2 in declination by --60\arcsec and 
--100\arcsec, respectively.  Velocity scale, $V_{LSR}-V_0$, refers to the main 
hyperfine component.
\label{fig:spectra1}
}
\end{figure}

\begin{figure}
\plotone{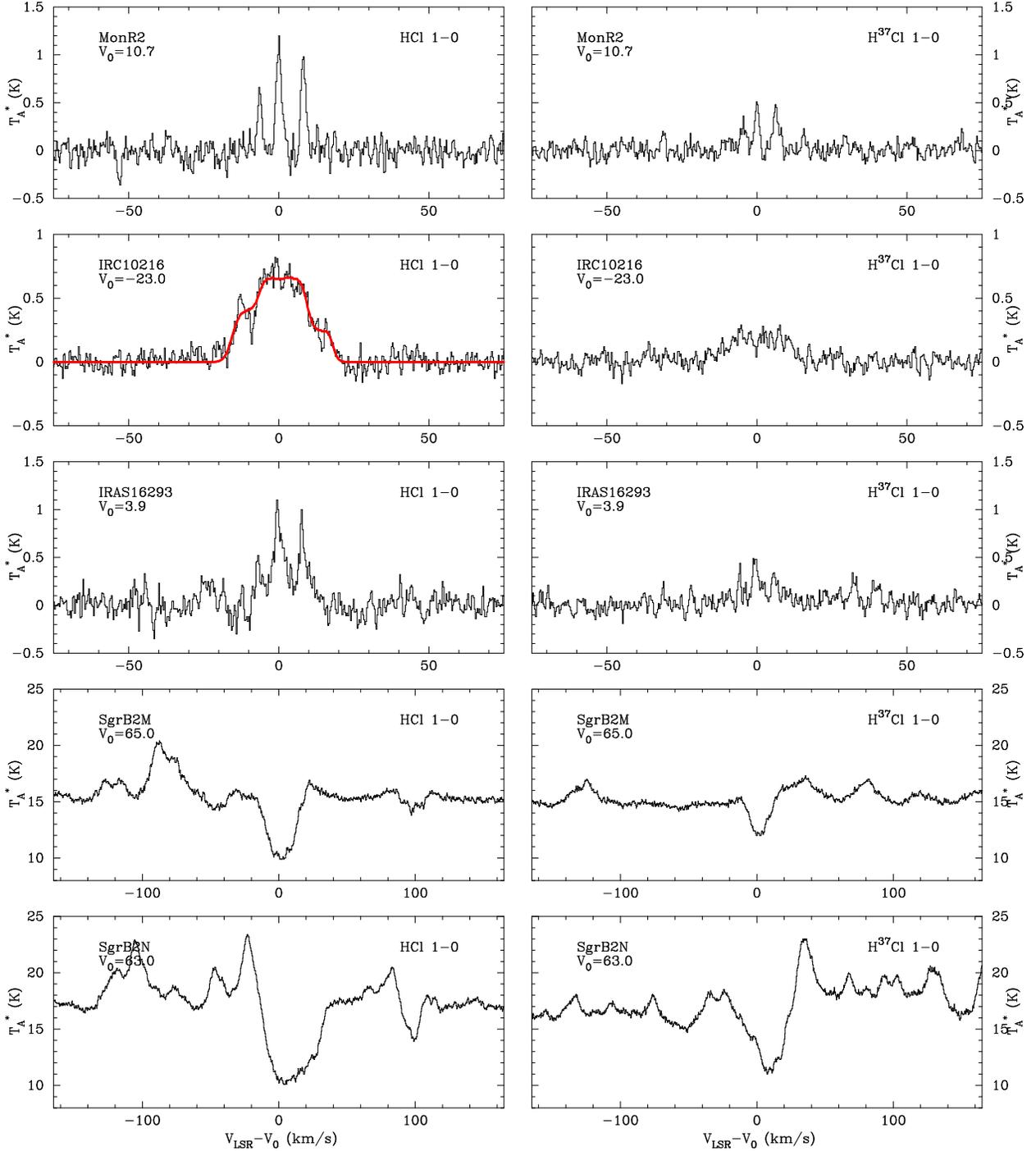}
\caption{HCl and H\tscl\  $J=1-0$ spectra toward MonR2, IRC+10216, IRAS16293, Sgr~B2(M),
and Sgr~B2(N). Velocity scale, $V_{LSR}-V_0$, refers to the main hyperfine
component.  For sources with detectable continuum, baseline is not removed to show the 
continuum level.
\label{fig:spectra2}
}
\end{figure}

\begin{figure}
\plotone{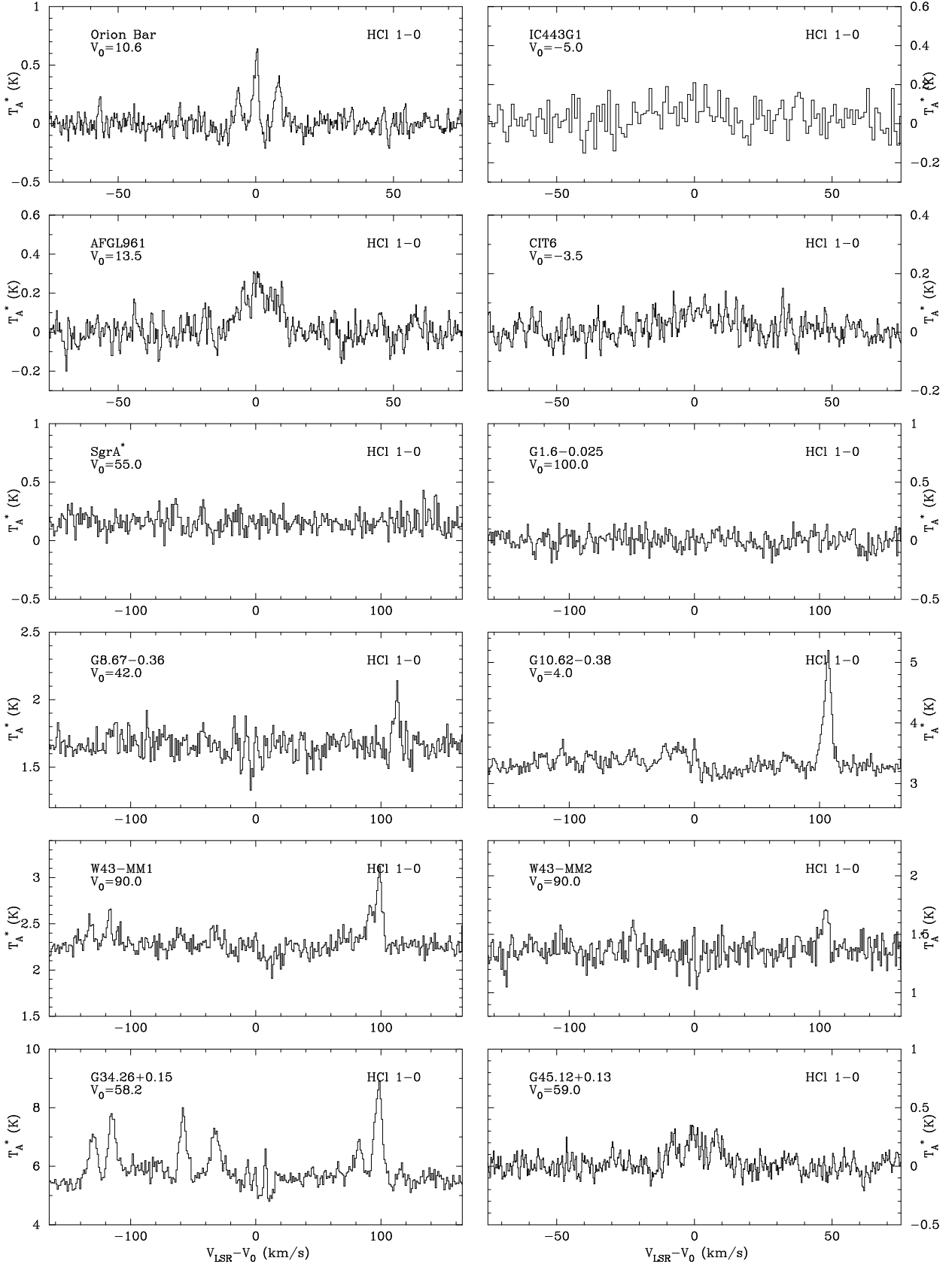}
\caption{HCl spectra toward rest of the sources in the sample. 
\label{fig:spectra4}
}
\end{figure}

For sources with matching HCl and H\tscl\ $J=1-0$ spectra, we present
in Figures~\ref{fig:spectra1}--\ref{fig:spectra3} HCl
spectra in the left column and H\tscl\ spectra in the right column. The
antenna temperature scale is kept the same for each source for
easy comparison of the line intensities. The velocity scale, $V_{\rm LSR}-V_0$,
refers to the main
hyperfine component; values of $V_0$ are denoted in each panel. Toward IRc2 in
OMC-1, the HCl and H\tscl\ line profiles are blended heavily with
spectral lines from other species, in the case for H\tscl\ 
predominantly methanol. It is not easy to untangle the blending and
derive useful information from the resulting line profile. Instead
we opted to observe two positions south of IRc2 with declination offsets of
--60\arcsec\ and --100\arcsec, respectively.  The two positions correspond
to local maxima in the HCl map of OMC-1 by \citet{spw1995}.
The (0, --60\arcsec) position also coincides with the CS clump LS1 
\citep{mundy1986}, and the (0, --100\arcsec) position
Orion~S. Figure~\ref{fig:spectra4} show the HCl $J=1-0$ spectra for
the remaining sources in the sample.

\subsection{Hyperfine Structure Fitting}

The HCl hyperfine structure, resulting from the interaction between
the electric field and chlorine nuclear spin, splits the $J=1-0$
transition into three components, with the outer two separated by
$-6.35$ and 8.22~\kms\ respectively from the strongest middle component for H\tfcl. 
For H\tscl, the separations are $-5.05$~\kms\ and 6.45 \kms, respectively.
Statistical weights for the upper level of the hyperfine transitions
dictate optically thin line strength ratio of 1:3:2. In the observed
sample, the three-component hyperfine pattern is clearly seen in
many sources, but the components often do not conform to the optically
thin ratio, indicating deviations from the optically thin limit or
differences in excitation between the three components.

\begin{figure}
\plotone{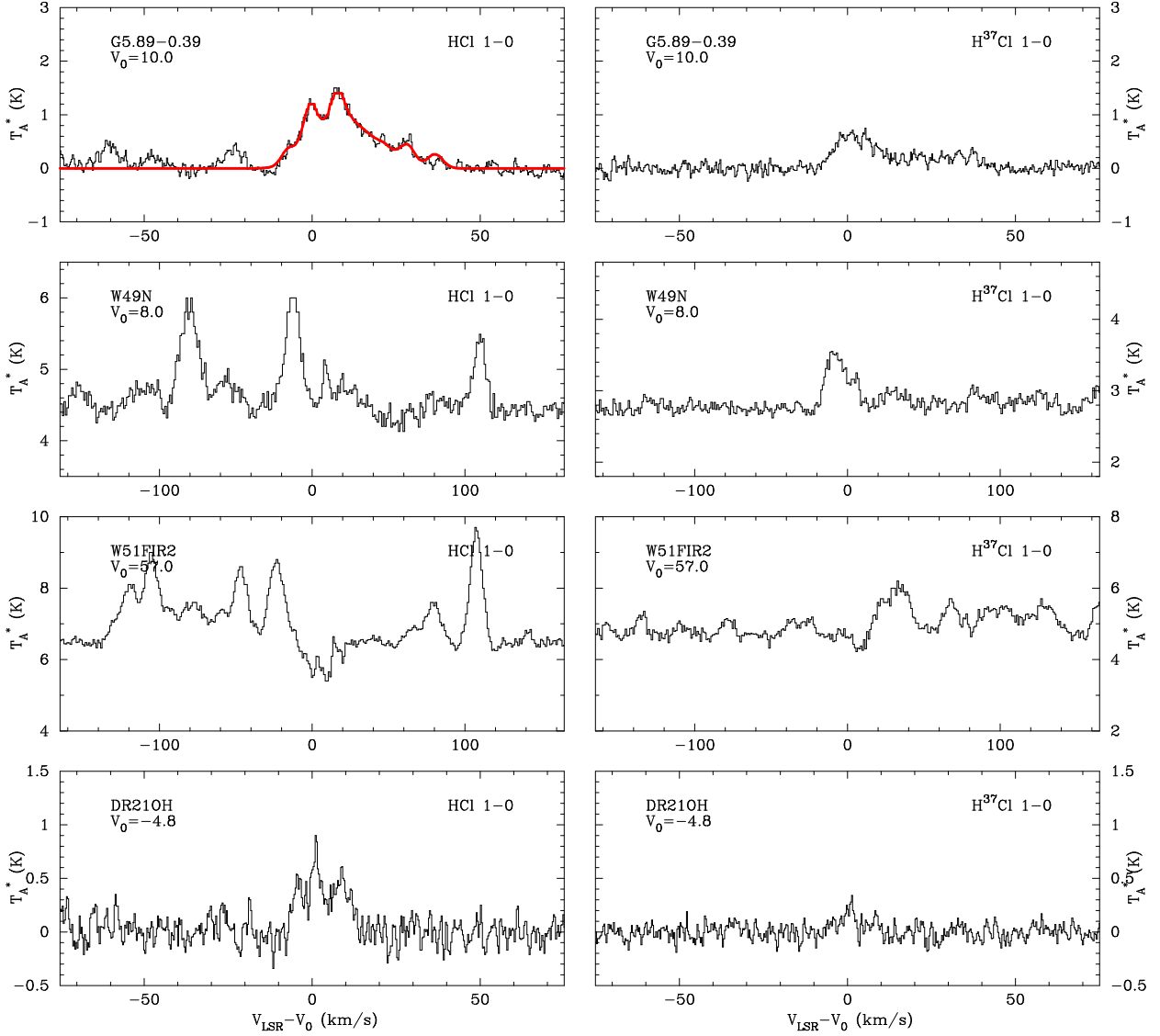}
\caption{HCl and H\tscl\  $J=1-0$ spectra toward G5.89-0.39, W49N, W51FIR2,
and DR21(OH). Velocity scale, $V_{LSR}-V_0$, refers to the main hyperfine
component.  For sources with detectable continuum, baseline is not removed 
to show the continuum level.
\label{fig:spectra3}
}
\end{figure}

Under the assumption that the hyperfine components share the same
excitation and kinematic characteristics, fits to the line profiles
can independently provide the optical depth and excitation temperature
\citep{kwan1975}. Such a utility is readily available in CLASS, the
spectral line data reduction package developed and maintained by IRAM.
The HFS fit in CLASS provides four parameters: $\Delta T\cdot\tau_{main}$,
$V_{LSR}$, $w$ the line width, and $\tau_{main}$. $\tau_{main}$ is the 
total optical depth from all hyperfine components of the line, and 
$\Delta T = T_{ex} - T_{bg}$ is the difference between the line 
excitation temperature and background temperature. 
Table~\ref{tab:line-fit} lists results of such hyperfine structure fits
for sources in our sample that show the $J=1-0$ 
HCl or H\tscl\ emission. Cases involving absorption are more complex, as
seen in G34.26+0.15 and W43-MM1 in Fig-\ref{fig:spectra3} for example.
They are often the result of a delicate balance between emission
and absorption features. We defer the treatment and
discussions for sources with absorption to a separate paper
\citep{hiro2010}.
 
In fitting the hyperfine profiles in $T_{mb}$ scale, one needs to be mindful
that the line emission may suffer from beam dilution when a source is
smaller than the telescope beam.  The uncorrected line temperatures would affect 
the excitation temperature and optical depth derived from HFS fit of the HCl
line profile. 
Thermal dust emission as seen in submillimeter continuum maps provide a
good measure for gas distribution in the source.  We compare the size of the
350$\mu$m dust cores to the telescope beam of 13\farcs5 to determine whether the
HCl emission from a given source suffers appreciable beam dilution. Most of our
sources in the sample are observed by \citet{dotson2010} with the SHARP camera at
the CSO, and the FWHM of the dust cores range from $\sim$30\arcsec\ to 
$\sim$60\arcsec.  We therefore do not consider beam dilution a significant
source of error for our sample of objects in deriving the excitation temperature 
and optical depth for the HCl $J=1-0$ transition. One exception may be
G5.89-0.39, where SMA continuum image at 870$\mu$m reveals a dust core
$\sim6\arcsec-8\arcsec$\ in size \citep{tang2009}, implying a factor of $\sim3$ in 
beam dilution in the observed line temperature. 

The HFS fits are sensitive to the signal-to-noise
ratio in the observed line profiles. $\Delta T$ and hence
$\tau_{main}$ are not well constrained for relatively noisy line
profiles. For sources with both HCl and H\tscl\ detections, one would 
expect that the values of $\Delta T$ derived from HFS fits of
the two transitions be very close.
In reality, they tend to agree within error bars, albeit rather large
error bars in some cases, resulting in questionable estimates of the
optical depths. We attempted to derive alternative estimates of the
optical depths for both HCl and H\tscl\ lines by imposing a fixed
value of $(\Delta T)^*$, a weighted average of the values obtained for
the two lines. Using $(\Delta T)^*$ and the $T_{mb}$ from Gaussian
fits of the three components in each profile, we derive $\tau$(HCl)$^*$ and 
$\tau$(H\tscl)$^*$ as listed in Table~\ref{tab:line-fit}. The error of
$T_{mb}$/$(\Delta T)$ amplified by $e^\tau$ translates into the 
error of $\tau^*$ for each components, resulting in relatively large errors 
for $\tau^*$ derived from this approach when individual components are optically 
thick.  We nonetheless consider $\tau^*$  a more reliable measure of the 
optical depth than that obtained directly from HFS fits.  For sources that 
have only HCl observations, $(\Delta T)^*$ and $\tau($HCl)$^*$ in
Table~\ref{tab:line-fit} are from direct HFS fits.  For G5.89-0.39, we
also present fits corrected for the likely case of beam dilution. Compared to the 
uncorrected case, the differences are mostly in the line excitation temperatures, 
with optical depth remain largely unchanged.

IRC+10216 shows a rather distinctive line profile. It is likely the 
result of convolution of the hyperfine structure with a shell profile often 
seen in circumstellar envelopes. As an approximation, we chose to fit 
two Gaussian profiles for IRC+10216 at LSR velocities
of $-17.8$~\kms and $-27.4$~\kms, respectively, to simulate the shell
profile in the hyperfine fits. As shown by the red curve in the
appropriate panel in Figure~\ref{fig:spectra2}, this approach adequately
fits the line profile, yielding a solution that indicates that the HCl
emission originated from a circumstellar shell with an expansion
velocity of 4.8 \kms, compared to an observed terminal velocity of 14.5 \kms\
(see for example \citet{schoier2001}).
 The [H\tfcl]/[H\tscl] isotopic ratio in the shell
is likely in excess of 4 from the optical depth shown in Table~\ref{tab:line-fit}.

CIT6 is widely regarded as similar to IRC+10216 with comparable
chemical and physical characteristics, but a lower
dust-to-molecular mass ratio \citep{zhang2009}. HCl is no exception in
that regard. It is clearly detected but the line is too weak to allow
for a good HFS fit. Nevertheless, this observation serves as an additional
proof that HCl as a species exists in detectable quantities in
circumstellar shells of evolved stars. IC443G1 is a dense core in a
ring of shocked material associated with SNR IC443
\citep{vanDishoeck1993}. There are tantalizing signs of HCl detection,
but the line is again too weak to derive any useful parameters.

G5.89-0.39 is another interesting case with a rather distinctive line profile. 
Both HCl and H\tscl\  line show a pronounced
red wing, indicating presence of higher velocity components.  A three-component
fit, as shown by the red curve in the top left panel of Figure~\ref{fig:spectra3},
represents an excellent match to the observed line profile. As listed in
Table~\ref{tab:line-fit}, the three components are at LSR velocities of 9.3,
17.9, and 38.3 \kms, with FWHM line widths of 4.9, 13.5, and 5.4 \kms,
respectively. From the interferometric study of \citet{hunter2008}, the broad
($\sim$13 \kms) line emission is seen in C$^{34}$S, SiO, CH$_3$OH, HC$_3$N,
C$^{17}$O, and H$^{13}$CO$^+$ toward the dust emission source SMA-N, in the 
north-northeast extension of the molecular ridge that bounds the northern and
western edge of the UCHII.  The 9.3~\kms\  feature likely originates from the
northwest and southwest section of the molecular ridge.  While the origin of the
38.3~\kms\  feature is not immediately clear, velocity features as highly
redshifted as 78~\kms\  and blue-shifted as -61~\kms, mostly from various masers,
were quoted by these authors in establishing a north-south bipolar outflow.  In
deriving the alternate optical depth of these velocity components in
Table~\ref{tab:line-fit}, we opted to use the $\Delta T$ from the HCl HFS fits
for the 17.9~\kms\ and 38.3~\kms\ components due to their low S/N ratio in the
H\tscl\ line profile.  From the derived $\tau^*$ in Table~\ref{tab:line-fit}, the 
[HCl/H\tscl] isotopic ratio in the source is most likely around 2, though the 
broad emission feature may have a higher ratio on the order of 3--4.

Sgr~A$^*$, the dynamical center of the Galaxy, yields no HCl detection 
to an rms level of 0.1 K at a resolution of 0.31~\kms. Similarly, 
G1.6-0.025, a giant molecular cloud in the Galactic center region with 
no apparent sign of star formation, also yields a non-detection. Sgr~B2(M) and
Sgr~B2(N), on the other hand, show broad strong absorption at both HCl and H\tfcl\
$J=1-0$ transitions. 

From Table~\ref{tab:line-fit}, it it clear that the [H\tfcl]/[H\tscl]
isotopic ratio, based on that of the optical depth, is rather varied,
from 4 and over in Mon~R2, IRC+10216, and DR21(OH), to around 3 in OMC-1, W3(OH),
and IRAS16293, down to around 2 in G5.89-0.39, and to
nearly reaching parity in W3IRS4 and W3IRS5. In comparison, the
terrestrial [\tfcl]/[\tscl] abundance ratio is $\sim$3.1.

\subsection{RADEX Simulations}

In addition to the optical depth obtained from hyperfine structure
fitting, we can derive other physical parameters from the line profiles
by running LVG simulations to fit the three HCl hyperfine components.
RADEX, a computer program developed and maintained by
\citet{vanderTak2007} for fast non-LTE analysis of interstellar line
spectra, was used in the LVG mode to find the best combination of
gas density and HCl column density that could reproduce the observed
line temperatures under given gas temperature.

The molecular data file for HCl under RADEX includes the hyperfine
levels of the lowest 8 rotational levels, up to 839 K above ground
level, with rates for radiative and collisional transitions among them.
The collisional rates between HCl and \mh\ are scaled from those of HCl and
helium \citep{neufeld1994}.
The average Galactic background radiation field adopted in RADEX consists of the
cosmic microwave background, average Galactic starlight in the solar
neighborhood in ultraviolet/visible/near-infrared part of the spectrum, and
single-temperature fit to the Galactic thermal dust emission for the
far-infrared and submillimeter part of the spectrum.  Contributions from 
non-thermal radiation in the Galaxy are also included for low frequency part of
the spectrum ($\le$30 GHz).  This composite interstellar radiation field
and any additional continuum background from user input provide radiative 
excitation in the RADEX formulism.  For the subset of our sample where 
HCl emission is observed, no significant continuum is seen in any of these
sources. We therefore use the default composite interstellar radiation field 
in RADEX simulations.

\begin{figure}
\plotone{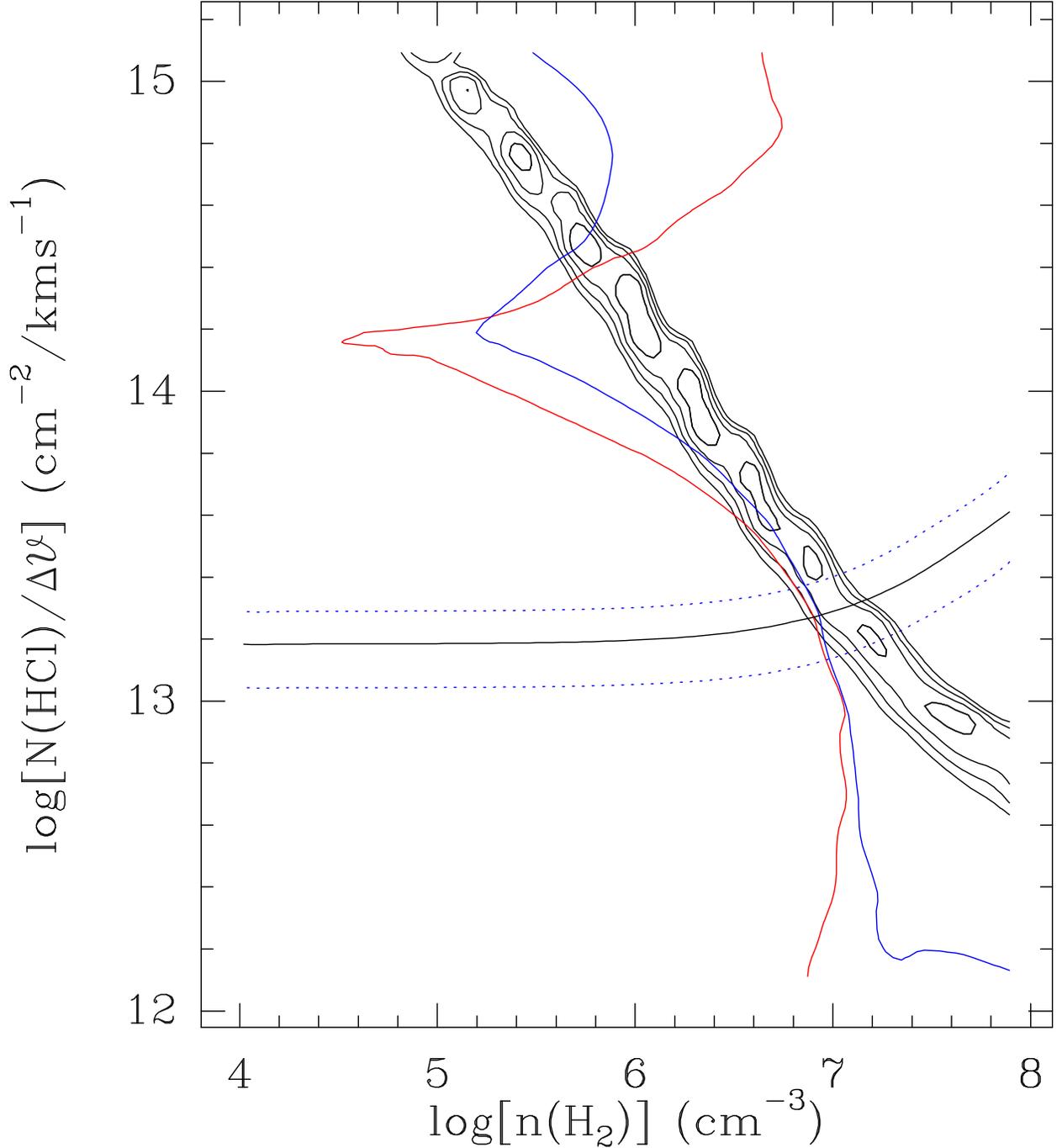}
\caption{A RADEX simulation run for OMC-1 (0,--60) with a fixed gas temperature 
of 45 K.  Contours are the relative $\chi^2$ between the observed and modeled line
temperatures, starting from 0.5 with a step of -0.1; blue and red
curves mark the hyperfine component ratios under optically thin limit (3
between F=5/2-3/2 and F=1/2-3/2 transitions in blue, 1.5 between F=5/2-3/2 
and F=3/2-3/2 transitions in red).  The solid smooth line and two dotted lines mark the
observed optical depth of the HCl line and its $\pm1\sigma$ error bars.
\label{fig:radex}
}
\end{figure}

Simulations were run in off-line mode to cover a large parameter
space, and relative $\chi^2$ was sought between the observed line 
temperatures and the model results relative to the observed line 
temperature.  Under the assumption that the observed HCl emission 
suffers no significant beam dilution, this process yields for a 
given gas temperature the most likely combination of gas density 
and HCl column density to reproduce the observed line strength of 
the hyperfine components.  Figure~\ref{fig:radex} shows a RADEX 
simulation run for OMC-1 (0, -60) with a gas temperature of 45 K. 
Contours are the relative $\chi^2$ starting from 0.5 with a step 
of -0.1; blue and red curves mark the hyperfine component ratios 
under optically thin limit (3 between F=5/2--3/2 and F=1/2--3/2 
transitions in blue, 1.5 between F=5/2--3/2 and F=3/2--3/2 
transitions in red).  The solid smooth line denotes the observed 
optical depth for the HCl line and the two dotted lines its 
$\pm1\sigma$ error bar. In cases where the quoted errors
for HCl optical depth in Table~\ref{tab:line-fit} exceed 30\%, we
generally use 30\% to set the error bar for the optical depth. The
cross-section between the $\chi^2$ contours and the optical depth
curves clearly confines gas density and HCl column density in an often
tight range.

Table~\ref{tab:radex-fit} lists for each source the adopted gas
temperature and \mh\ column density, the gas density and HCl column
density derived from RADEX simulations, as well as the HCl abundance.
Since there is no molecular data file for H\tscl\ to use under RADEX,
we treat it the same as HCl and fit the H\tscl\ line temperatures with
the same procedure. Instead of confining the ranges of gas density and
H\tscl\ column density with the relative $\chi^2$ contours and optical 
depth curves, we use gas density ranges found in the corresponding HCl 
simulation of the source in place of optical depth curves, ensuring 
that both HCl and H\tscl\ emission originated from the same region. 
The resulting H\tscl\ column density is listed in Table~\ref{tab:radex-fit} 
for sources with observed H\tscl\ line emission. For these sources we also
list the \tfcl/\tscl\ abundance ratio X[\tfcl/\tscl] as measured by
the ratio of HCl and H\tscl\ column densities.  \citet{Cernicharo2010b} recently
suggested that the integrated intensity ratio be a good gauge for HCl and
H\tscl\ isotopic ratio.  We listed in Table~\ref{tab:radex-fit} the integrated
intensity ratio, R, which appears to be in good agreement with our derived
chlorine isotopic ratio.

\section{Discussion}
\subsection{Where to Look for HCl Emission?}

From Table~\ref{tab:radex-fit} it is clear that HCl emission is associated with
warm dense gas with a density on the order of 10$^6$ cm$^{-3}$. Given a critical
density of a few $\times10^8$ cm$^{-3}$,  molecular excitation is mostly
sub-thermal, dominated by radiative processes rather than collisions with \mh.
Sources with strong continuum and/or dust emission 
would likely yield strong HCl line emission.

The derived gas densities for the two OMC-1 cores appear to be somewhat higher
than $\sim5\times10^6$ established by the multi-transitional study of HC$_3$N
\citep{bergin1996}.  This may indicate that the gas temperatures adopted for
these cores for the RADEX simulations should have been higher. Another
factor may be the average Galactic background radiation field used
in RADEX for radiative excitation.  As HCl transitions are short-ward of
$\sim$480$\mu$m in wavelength, radiation from warm dust would contribute  
significantly to the molecular excitation.  Including local warm dust in the
radiation field for RADEX simulation would produce a more precise picture of the
gas where HCl emission originates.

The non-detection toward Sgr~A$^*$ likely reflects a genuine lack of HCl
molecules in the circumstellar disk surrounding Sgr~A$^*$.  \citet{latvakoski1999}
show a warm dust ring $\sim80\arcsec$ in size, as well as the "minispiral"
within the ring. The dust ring coincides with a molecular ring with sharp inner
edges seen in HCN \citep{gusten1987}.  With our 13\farcs5 telescope beam, we are
clearly probing the molecular cavity around Sgr~A$^*$.  Using a FWHM (15
\kms) and the integrated line intensity of the C$^{18}$O J=$2-1$ line we obtain 
toward Sgr~A$^*$, and assuming LTE at 250 K \citep{bradford2005}, we derive a 
C$^{18}$O column density of $9\times10^{15}$ cm$^{-2}$, and an \mh\ column
density of $5\times10^{22}$ with a C$^{18}$O abundance of $1.7\times10^{-7}$ 
\citep{Goldsmith1997}.  A upper limit of $4\times10^{12}$ cm$^{-2}$ is derived
for the HCl column density by RADEX simulations for an HCl line of 0.1 K with a 
FWHM of 15 \kms\ (from the C$^{18}$O spectrum), yielding an upper limit of
$8\times10^{-11}$ for HCl abundance relative to \mh\ toward Sgr~A$^*$, markedly
lower than those listed in Table~\ref{tab:radex-fit}.

The non-detection in G1.6-0.025 is intriguing.  The source is a giant molecular 
cloud located in the Central Molecular Zone (CMZ) of the Galaxy, a region 
characterized by elevated gas temperature (typically $\sim$70 K), higher gas density 
($\ge10^4$~cm$^{-3}$), and highly supersonic internal velocity dispersion 
($\sim$15--50~\kms) \citep{morris1996}. Dense cloud cores are seen in various 
molecular tracers \citep{gardner1981, gardner1987, wp1989, kuiper1993, menten2009}, 
yet there is little sign of apparent star forming activity in the cloud.  For the 
extended 50~\kms\ cloud of G1.6-0.025, we derive a C$^{18}$O column density of
$2\times10^{15}$ from the C$^{18}$O J=$2-1$ spectrum taken at the CSO, assuming LTE
at a gas temperature of 30 K \citep{menten2009}. Again with a C$^{18}$O
abundance of $1.7\times10^{-7}$, the cloud has an \mh\ column density of
$1\times10^{22}$ cm$^{-2}$.  Similar to Sgr~A$^*$, we obtain an upper limit of
$7\times10^{12}$ for HCl column density by modeling a HCl line of 0.1 K with
28~\kms\ in line width, yielding an HCl abundance of $\la7\times10^{-10}$ relative 
to \mh. It would look as though the HCl non-detection in
the 50~\kms\ cloud of G1.6-0.025 is the result of a combination of relative low
\mh\ column density and HCl abundance.
 
It would be interesting to see if other clouds in CMZ have similarly low HCl
abundance seen in Sgr~A$^*$ and the 50~\kms\ cloud of G1.6-0.025.  A prime
target would be M-0.02-0.07, the so-called 20 \kms\ cloud $\sim$2' northeast of
Sgr~A$^*$.  A common trait among the three clouds is the likely interaction with
a SNR \citep{menten2009}.  It is not clear though how such interaction would
lead to relatively low production of HCl.

In the same vein, IC443G1 may represent another case where chemistry plays an 
important role in the detection of HCl.  Clump G1 is a dense core in a ring 
of shocked material associated with SNR IC443 \citep{vanDishoeck1993}.  It has a
relatively rich chemistry compared to the pre-shock gas, but lacks many of the
more complicated molecular species, such as methanol.  These authors argued that 
such deficiency could be the result of disruption of chemical network
brought on by the infusion of atomic hydrogen to the shocked gas by repeated
shocks, as atomic hydrogen rapidly destroys many of the intermediate molecules 
in the reaction network.  To derive an estimate of
HCl column density from the low signal-to-noise detection, we use RADEX
simulation to reproduce an HCl line of 0.2 K with a FWHM of 10~\kms\
\citep{vanDishoeck1993}.  The resulting HCl column density of
$3\times10^{12}$~cm$^{-2}$, when compared to an \mh\ column density of
$1\times10^{22}$ \citep{vanDishoeck1993}, leads to an HCl abundance of
$3\times10^{-10}$.  Again, the weak HCl line emission from IC443 Clump G1 may be
the result of relatively low HCl abundance and \mh\ column density.
More observations of similar cores are needed to
confirm that shocked gas may have a somewhat lower HCl abundance compared to
other similarly dense gas in star forming regions, PDRs, and Ultra-compact HII
regions.

Ultra-compact HII regions with reasonably high emission measures have proven to be a
treasure trove for detecting HCl absorption (see Figure~\ref{fig:spectra4}).  Such
cases typically involve a delicate balance between a emission feature and an
absorption feature.  Our forthcoming paper \citep{hiro2010} will discuss
these cases.

\subsection{The Case of IRC+10216}

In Figure~\ref{fig:spectra2} we shown that the HCl line profile for
IRC+10216 could be adequately fit by two Gaussians centered at
$-17.8$~\kms and $-27.4$~\kms\ respectively, indicating that the HCl
emission is associated with a circumstellar shell with an expansion
velocity of 4.8 \kms.  The HCl emission is likely from a density-enhanced
shell at $\sim$15\arcsec\ from the central star as seen in H$^{13}$CN
\citep{schoier2007} and SiO \citep{schoier2006}, and modeled by
\citet{cordiner2009}. Radiation from the central
star, as well as from local dust, contribute significantly to the molecular
excitation. We therefore consider the HCl abundance and [\tfcl/\tscl]
abundance ratio for IRC+10216 as listed in Table~\ref{tab:radex-fit}
highly uncertain. More sophisticated modeling efforts such as
\citet{schoier2001} and \citet{wyrowski2006} are required to correctly
interpret the results for IRC+10216.  \citet{Cernicharo2010b} recently reported
an HCl abundance of $5\times10^{-8}$ from their Herschel SPIRE and PACS 
observations of IRC+10216.  It may be interesting to note that we derive an
upper limit of $8\times10^{-8}$ for the HCL abundance in CIT6 by treating the low
signal-to-noise detection as a 0.2 K line with 20~\kms\ in line
width.  Similar to the procedure used in Sgr~A$^*$ and G1.6-0.025, the resulting
HCl column density is $5\times10^{12}$~cm$^{-2}$, with an adopted gas temperature
of 40 K and an \mh\ column density of $6.3\times10^{19}$~cm$^{-2}$
\citep{zhang2009}. 

\subsection{\tfcl/\tscl\  Abundance Ratio}

As listed in Table~\ref{tab:radex-fit}, the \tfcl/\tscl\ abundance ratio is 
rather varied from 
around unity to around 3 and over. \citet{Cernicharo2000} reported an 
X[\tfcl/\tscl] ratio of 3.1$\pm$0.6 from line intensity ratios of
NaCl, KCl, and AlCl in IRC+10216.  Recent Herschel HIFI observations
show an X[\tfcl/\tscl] ratio of 2.7 in NGC6334I \citep{Lis2010} and 
$2.1\pm0.5$ in W3A \citep{Cernicharo2010a}. These compare to the 
terrestrial value of $\sim$3.1.  It is interesting to note that the 
\tfcl/\tscl\ abundance ratio comes in a good agreement with the integration 
line intensity ratio of the HCl and H\tscl\ line, as predicted by the modeling 
efforts of \citet{Cernicharo2010a} for their Herschel HIFI observations of W3A.

One may argue that the low isotopic ratio observed in W3IRS4 might be the 
result of observing an unresolved, very optically thick source.  DR21(OH), 
which is at a comparable distance ($\sim$2.5 kpc vs. $\sim$2.2 kpc for W3IRS4), 
with a comparable dust core size at 350$\mu$m \citep{dotson2010} and HCl opacity, 
argues against this scenario.  

Clearly, variations in HCl isotopic ratio are highly
localized. \citet{sfl1996} suggested one likely origin for such
variations through explosive nucleosynthesis, whereby enrichment of
one isotopologue over the other, as synthesized in massive stars, is
mixed into the surrounding molecular cloud through explosive processes
such supernova explosions.

Models for nucleosynthesis in Type Ia supernovae (SNe Ia; \citet{Travaglio2004}
and references therein) and Type II supernovae (SNe II; 
\citep{Nomoto1997, Nomoto2006, Kobayashi2006}) 
indeed show a general deficiency of \tscl\ in the processed material in most
cases. In SNe Ia, the \tfcl/\tscl\ abundance ratios varies from 3.5 to 5.4 among
models of different ignition conditions in the central white dwarf before the
thermonuclear runaway \citep{Travaglio2004}.  In SNe II, the elemental yields
of the nucleosynthesis depend on the mass and metallicity of the progenitor, 
as well as the explosion energy. \citet{Kobayashi2006} shows
the \tfcl/\tscl\ abundance ratios produced in SNe with high explosion energy
($\ga$10 E$_{51}$ with E$_{51} \sim 10^{51}$ erg, so called hypernovae) are
typically in the range of $1.1-2.8$ for varying progenitor mass and
metallicity, while the ratios varies in a larger range  (to as high as 18) in
normal SNe (with explosion enery $\sim$E$_{51}$).  It is inetresting to note that
in normal SNe II with relatively large progenitor mass ($\ga$25M$_\sun$) and
metallicity (Z$\sim$0.02), the \tfcl/\tscl\ abundance ratios are consistently below
unity to as low as $\sim$0.5 \citep{Kobayashi2006}.
Given enough mixing with the surrounding molecular clouds, these
ratios could support the observed range of the \tfcl/\tscl\ abundance ratio.

W3IRS4, and possibly W3IRS5 present interesting exceptions of relative \tscl\ 
deficiency predicted in most nucleosynthesis models SNe. This likely confines the 
progenitors of the SNe to be stars of relatively high mass ($\ga$25M$_\sun$) and high 
metallicity (Z$\sim$0.02).

\citet{kawabata2010} and \citet{perets2010} recently 
reported the discovery of a new category of subluminous supernovae.  Although
the origin for such faint supernovae is still controversial, these authors
reported the unusual composition in the supernovae ejecta with rich helium and 
calcium.  It would be interesting to examine the nucleosynthesis yields in this 
class of supernovae in regard of the \tfcl/\tscl\ abundance ratios.

\section{CONCLUSION}

We conducted a comprehensive survey of HCl $J=1-0$ line in the Galaxy.
Of the 27 sources/positions observed, fourteen show emission, nine
show absorption, two showed marginal detection in emission, and two
are non-detections. Fourteen of the sources/positions were also
observed in the H\tscl\ $J=1-0$ transition.

RADEX simulations show that HCl emission is mostly
associated with warm dense gas, of order $10^6$ cm$^{-3}$.  HCl abundance is
fairly uniform, in the range $10^{-10}$ to a few $\times10^{-9}$, in
general agreement with previous studies of OMC-1 by \citet{spw1995} and
\citet{sfl1996}.
However, \tfcl/\tscl\  abundance ratios are found to vary considerably from
around unity in W3IRS4 and W3IRS5 to $\sim$5 in DR21(OH).  They
are in good agreement to the integrated intensity ratios of the HCl and H\tscl\
line as suggested by \citet{Cernicharo2010a}. 

The variations of the \tfcl/\tscl\  abundance ratios are found to be highly
localized. They could be supported by the varying yields of nucleosynthesis of
supervoae from different progenitors.  The low \tfcl/\tscl\  abundance ratios 
seen in W3IRS4 and W3IRS5 likely confine the progenitors of the supernovae to 
stars of relatively high mass ($\ga$25M$_\sun$) and high metallicity (Z$\sim$0.02).

\acknowledgements

Caltech Submillimeter Observatory (CSO) is supported through NSF grant
AST-0540882.

\clearpage
\begin{deluxetable}{lrrrcc}
\tablewidth{0pt}
\tablecolumns{6}
\tablecaption{Source list of the survey\label{tab:obs-sum}}
\tabletypesize{\scriptsize}
\tablehead{ \colhead{Source} & \colhead{RA(J2000)} & \colhead{Dec(J2000)} &
\colhead{$v_{LSR}$} & \colhead{HCl} & \colhead{H\tscl}}
\startdata
W3IRS4	& 02$^h$25$^m$30\fs90	& 62\degr06\arcmin20\farcs29 & -42 & Y & Y\\
W3IRS5	& 02 25 40.46	& 62 05 52.27 & -39 & Y & Y\\
W3(OH)	& 02 27 03.82	& 61 52 24.60 & -44 & Y & Y\\
OMC-1	& 05 35 14.18	& -05 22 30.55 & 8 & Y & Y\\
OMC-1 (0, -60)	& 05 35 14.18	& -05 23 30.55 & 8 & Y & Y\\
OMC-1 (0, -100) & 05 35 14.18	& -05 24 12.55 & 8 & Y & Y\\
Orion Bar	& 05 35 23.99	& -05 24 59.97 & 9 & Y & N\\
MonR2	& 06 07 46.19	& -06 23 08.30 & 10.5 & Y & Y\\
AFGL961	& 06 34 37.64	& 04 12 43.82 & 15 & Y & N\\
IRC+10216	& 09 47 57.30	& 13 16 43.05 & -26 & Y & Y\\
CIT6	& 10 16 02.34	& 30 34 18.30 & -2 & Y & N\\
IRAS16293	& 16 32 22.91	& -24 28 35.52 & 4 & Y & Y\\
Sgr~B2(M)	& 17 47 20.16	& -28 23 04.97 & 62 & Y & Y\\
Sgr~B2(N)	& 17 47 19.90	& -28 22 17.78 & 65 & Y & Y\\
Sgr~A$^*$	& 17 45 40.04	& -29 00 28.10 & 60 & Y & N\\
G1.6-0.025	& 17 49 20.61	& -27 34 05.46 & 55 & Y & N\\
G5.89-0.39	& 18 00 30.43	& -24 04 01.48 & 10 & Y & Y\\
G10.62-0.38	& 18 10 28.67	& -19 55 50.08 & 0.4 & Y & N\\
W43-MM2	& 18 47 36.70	& -02 00 51.90 & 90 & Y & N\\
W43-MM1	& 18 47 47.00	& -01 54 28.00 & 90 & Y & N\\
G34.26+0.15	& 18 53 18.58	& 01 14 58.41 & 54.6 & Y & N\\
W49N	& 19 10 13.62 &  09 06 17.3 & 0.0 & Y & Y\\
G45.12+0.13	& 19 13 27.84 &  10 53 36.6 & 58.0 & Y & N\\
W51FIR2		& 19 23 44.17 &  14 30 33.4 & 57.0 & Y & Y\\
DR21(OH)	& 20 39 00.70 & 42 22 46.7 & -4.7 & Y & Y\\
\enddata
\end{deluxetable}

\clearpage
\begin{deluxetable}{lrrrrrrrrrrrrrrrr}
\tablecolumns{17}
\tablewidth{0pt}
\tabletypesize{\scriptsize}
\rotate
\tablecaption{Hyperfine structure fits for HCl and H\tscl\  line emission.\label{tab:line-fit}}
\tablehead{
\colhead{} & \colhead{} & \colhead{} & \multicolumn{5}{c}{HCl} & \colhead{} &
\multicolumn{5}{c}{H$^{37}$Cl} & \colhead{} & \colhead{} & \colhead{} \\
\cline{4-8} \cline{10-14} \\
\colhead{Source} & \colhead{$v_{LSR}$} & \colhead{$w$} &
\colhead{S$_1$} & \colhead{S$_2$} & \colhead{S$_3$} & \colhead{$(\Delta T\cdot\tau)$} & \colhead{$\tau$} & \colhead{} &
\colhead{S$_1$} & \colhead{S$_2$} & \colhead{S$_3$} & \colhead{$(\Delta T\cdot\tau)$} & \colhead{$\tau$} & 
\colhead{$(\Delta T)^*$} & \colhead{$\tau$(HCl)$^*$} & 
\colhead{$\tau$(H\tscl)$^*$}}
\startdata
W3IRS4   & -44.0 & 3.2 & 4.0 & 7.7 & 5.9 & 7.9(1.2) & 3.7(1.1) & 
& 4.4 & 8.0 & 6.7 & 6.7(0.6) & 1.8(0.6) & 2.7(0.8) & 2.5(2.1) & 2.8(2.5)  \\
W3IRS5   & -37.9 & 4.8 & 15.7 & 24.9 & 24.3 & 24.6(1.9) & 5.8(0.7) &
& 12.8 & 22.7 & 17.8 & 16.0(0.9) & 3.1(0.3) & 4.6(0.4) & 5.0(2.3) & 3.3(1.0)  \\
W3(OH)   & -49.2 & 4.9 & 7.9 & 12.0 & 12.2 & 11.3(1.8) & 5.3(1.2) &
& 1.5 & 9.2 & 7.0 & 3.9(0.5) & 0.7(0.7) & 2.2(0.6) & 4.9(6.9) & 1.9(2.3)  \\
OMC1 (0, -60) & 8.4 & 3.1 & 9.6 & 17.6 & 13.2 & 15.1(1.4) & 2.0(0.5) &
& 4.9 & 7.6 & 4.4 & 5.4(0.7) & 1.0(0.7) & 7.5(0.7) & 2.2(0.6) & 0.7(0.2)\\
OMC1 (0, -100) & 7.9 & 3.2 & 10.7 & 19.3 & 17.1 & 20.3(2.7) & 3.5(0.9) &
& 4.2 & 8.5 & 7.0 & 6.0(0.6) & 0.7(0.6) & 6.1(0.7) & 3.6(1.6) & 1.0(0.3)\\
ORIONBAR\tablenotemark{a} & 10.6 & 1.9 & 1.2 & 2.6 & 1.8 & 3.6(0.4) & 2.3(0.6) & 
& \nodata  & \nodata & \nodata & \nodata & \nodata & 1.6(0.4)  & 2.3(0.6)  & \nodata \\
MonR2   & 10.7 & 1.7 & 2.3 & 4.8 & 4.0 & 9.0(0.6) & 3.3(0.4) & 
& 1.1 & 2.0 & 2.0 & 4.1(0.8) & 3.6(1.4) & 2.5(0.5) & 4.4(4.3) & 1.1(0.5)  \\
AFGL961\tablenotemark{a} & 13.4 & 6.5 & 1.3 & 3.4 & 2.3 & 1.1(0.1) & 0.4(0.3) & 
& \nodata  & \nodata & \nodata & \nodata & \nodata & 2.8(2.5)  & 0.4(0.3)  & \nodata \\
IRC+10216  & -17.8 & 6.4 & 5.2 & 2.8 & 5.5 & 9.6(0.1) & 17.8(0.1) & 
& 0.4 & 0.8 & 1.4 & 0.8(0.3) & 5.0(3.0) & 0.54(0.01) & 7.3(6.8) & 0.8(0.5)   \\
        & -27.4 & 8.5 & 6.5 & 5.9 & 9.1 & 6.0(0.1) & 6.5(0.1) & 
& 2.0 & 1.8 & 2.0 & 3.2(3.0) & 11.6(1.5) & 0.93(0.01) & 3.2(0.6) & 0.8(0.2)  \\
IRAS16293  & 3.9 & 4.2 & 3.0 & 8.2 & 6.2 & 4.5(0.3) & 1.1(0.4) & 
& 1.4 & 3.2 & 2.3 & 2.0(0.5) &  0.6(0.7) & 3.7(1.1) & 1.1(0.3) & 0.4(0.1)  \\
G5.89-0.39\tablenotemark{b,d} & 9.3 & 4.9 & 2.7 & 8.0 & 6.0 & 3.5(0.7) & 0.7(0.5) & 
& 1.0 & 5.2 & 3.9 & 2.1(0.2) & 0.5(0.4) & 4.6(0.4) & 0.8(0.2) & 0.5(0.2)  \\
         & 17.8 & 13.5 & 10.3 & 17.0 & 10.3 & 2.8(0.2) & 0.3(0.2) & 
& 3.2 & 5.3 & 2.4 & 2.0(0.5) & 4.1(1.2) & 9.4(4.6) & 0.3(0.3) & 0.08(0.07)  \\
         & 38.3 & 5.4 & 1.7 & 5.3 & 1.7 & 1.6(0.2) & 0.1(0.1) & 
& 0.6 & 2.1 & 0.4 & 0.6(0.1) & 0.1(1.1) & 15(8) & 0.1(0.1) & 0.03(0.03)  \\
G5.89-0.39\tablenotemark{c,d}   & 9.3 & 4.9 & 8.0 & 24.0 & 17.8 & 10.8(2.1) & 0.7(0.5) & 
& 2.8 & 15.9 & 11.7 & 6.4(0.5) & 0.5(0.4) & 13.4(1.3) & 0.8(0.3) & 0.5(0.3)  \\
          & 17.9 & 13.3 & 30.6 & 50.6 & 31.5 & 8.7(0.5) & 0.4(0.2) & 
& 9.7 & 15.9 & 7.2 & 6.3(1.4) & 4.3(1.2) & 24(11) & 0.4(0.3) & 0.10(0.08)  \\
          & 38.3 & 5.4 & 5.2 & 15.8 & 5.0 & 4.8(0.3) & 0.1(0.1) & 
& 2.0 & 6.4 & 1.1 & 1.9(0.2) & 0.1(0.8) & 48(24) & 0.1(0.1) & 0.03(0.03)  \\
G45.12+0.13\tablenotemark{a} & 58.5 & 3.8 & 1.8 & 2.8 & 2.5 & 3.1(0.7) & 5.2(1.8) & 
& \nodata  & \nodata & \nodata & \nodata & \nodata & 0.6(0.2)  & 5.2(1.8)  & \nodata\\
DR21(OH)\tablenotemark{e} & -3.6 & 4.7 & 3.4 & 6.9 & 5.6 & 4.4(0.4) & 2.6(0.6) & 
& 0.6 & 1.9 & 0.9 & 0.9(0.1) & 0.1(1.0) & 1.7(0.2) & 2.7(1.7) & 0.4(0.3)  \\
\enddata
\tablecomments{Numbers in parentheses are quoted error bars; S$_1$, S$_2$, and
S$_3$ are integrated intensity for the $F=1/2-3/2$, $5/2-3/2$, and $3/2-3/2$ component of
the $J=1-0$ transition, respectively.}
\tablenotetext{a}{$(\Delta T)^*$ and $\tau$(HCl)$^*$ are from HFS fit of the HCl line profile.}
\tablenotetext{b}{assuming no appreciable beam dilution.}
\tablenotetext{c}{accounting for a likely  $\times$3 beam dilution.}
\tablenotetext{d}{$(\Delta T)^*$ for the 18.0 \kms\ and 38.3 \kms\ components are from the HCl 
measurements alone as the low S/N in the H\tscl\ counterparts prevent reliable HFS fits.}
\tablenotetext{e}{$(\Delta T)^*$ is from the HCl measurement alone as the low S/N in the H\tscl\ 
counterparts prevent reliable HFS fit.}
\end{deluxetable}

\clearpage
\begin{deluxetable}{llrrrrrcc}
\tablewidth{0pt}
\tabletypesize{\scriptsize}
\tablecolumns{9}
\tablecaption{RADEX fits for HCl and H\tscl\  line emission.\label{tab:radex-fit}}
\tablehead{ \colhead{Source} & \colhead{$T_K$} & \colhead{N(\mh)} & 
 \colhead{n(\mh)} & \colhead{N(HCl)} & \colhead{N(H\tscl)} & \colhead{X[HCl]} &
\colhead{X[\tfcl/\tscl]} & \colhead{R}}
\startdata
W3IRS4     &    $55^a$ & $3(23)^a$ & (1--3)(6) & (6.3--9.1)(13) &
(8.1--10.2)(13) & 3(-10) & $0.8^{+0.3}_{-0.2}$ & 0.9\\
W3IRS5     &    $100^a$ &  $5(23)^a$ & (8--13)(5) & (1.5--2.2)(14) &
(1.3--1.8)(14) & 4(-10) &  $1.2^{+0.5}_{-0.4}$ & 1.2\\
W3OH       &    $150^a$ &  $4(23)^a$ & (2--3)(5) &  (1.5--2.4)(14) &
(7.0--9.9)(13) & 5(-10) &  $2.3^{+1.2}_{-0.8}$ & 1.8\\
OMC-1(0,-60) &  $45^b$  &  $7.4(22)^c$  & (1--2)(7) & (5.1--6.3)(13) &
(1.8--2.8)(13) & 8(-10) &  $2.5^{+0.9}_{-0.7}$ & 2.4\\
OMC-1(0,-100) & $30^b$  &  $7.4(22)^c$  & (8--11)(6) & (1.3--1.5)(14) &
(5.5--7.8)(13) & 2(-9) &   $2.1^{+0.5}_{-0.5}$ & 2.4\\
Orion Bar   &   $75^d$  &  $2.3(22)^c$  & (1--3)(6) & (2.5--4.3)(13) & \nodata &
1(-9)  & \nodata & \nodata\\
MonR2    &      $70^e$  &  $3.9(22)^{e,f}$ & (1--3)(6) &  (5.7--9.3)(13) &
(2.8--3.7)(13) & 2(-9) &  $2.3^{+1.0}_{-0.7}$ & 2.2\\
AFGL961   &     $20^g$  &  $1.1(22)^h$ &  (7--10)(6)  &   (1.9--2.5)(13) &
\nodata & 2(-9) & \nodata & \nodata\\
IRC+10216 &     $50^i$  &  $7(20)^j$  &   (5--30)(4) &  (5.7--8.8)(14) &
(2.6--3.0)(14) & 1(-6) &  $2.6^{+0.8}_{-0.7}$ & 4.2\\
IRAS16293  &    $80^k$  &  $2(23)^k$  &   (1--4)(6)  & (3.0--4.7)(13) &
(1.3--2.4)(13) & 2(-10) &  $2.1^{+1.6}_{-0.8}$ & 2.5\\
G5.89-0.39\tablenotemark{1}  &   $70^l$  &  $3.1(22)^l$ &  (1--5)(6) &
(5.4--7.5)(13) & (2.8--3.3)(13) & 2(-9) & $2.1^{+0.6}_{-0.5}$ & 2.6\\
G5.89-0.39\tablenotemark{2}  &   $70^l$  &  $3.1(22)^l$ &  (3--13)(6) &
(7.4--10.1)(13) & (3.2--3.9)(13) & 3(-9) & $2.5^{+0.7}_{-0.6}$ & 2.6\\
G45.12+0.13 &   $25^m$  &  $1.2(23)^m$  & (9-12)(5)   &  (1.5--2.0)(14) &
\nodata & 1(-9) & \nodata & \nodata\\
DR21(OH)   &    $40^n$  &  $3(23)^n$  &   (1--3)(6) & (7.1--11.3)(13) &
(1.2--2.5)(13) & 3(-10) &  $5.0^{+4.1}_{-2.1}$ & 4.7\\
\enddata
\tablecomments{The numbers in the rightmost parentheses are exponents to the
power of 10; R is the integrated intensity ratio of the HCl and H\tscl\ $J=1-0$
transition.}
\tablenotetext{1}{assuming no appreciable beam dilution.}
\tablenotetext{2}{accounting for a likely  $\times$3 beam dilution.}
\tablerefs{
$^a$ \citet{Helmich1994}; $^b$ \citet{bergin1994}; $^c$ \citet{Goldsmith1997};
$^d$ \citet{Lis1998}; $^e$ \citet{Choi2000}; $^f$ \citet{Tafalla1997}; $^g$
\citet{Williams2009}; $^h$ \citet{Schneider1998}; $^i$ \citet{schoier2000}; 
$^j$ \citet{Cernicharo2000}; $^k$ \citet{vanDishoeck1995}; $^l$ \citet{Thompson1999};
$^m$ \citet{Hofner1997}; $^n$ \citet{Chandler1993}
}
\end{deluxetable}
\end{document}